\documentclass[preprint,authoryear,11pt]{elsarticle}
\usepackage[margin=0.80in]{geometry}
\usepackage{float}
\usepackage{adjustbox}
\usepackage{wrapfig}
\usepackage{soul}
\usepackage{footnote}

\usepackage{multirow}
\usepackage{graphics}
\usepackage{pifont}
\usepackage{graphicx}
\usepackage{epstopdf}

\usepackage{caption}
\usepackage{subcaption}

\usepackage{amssymb,amstext,amsmath}
\usepackage{float}
\usepackage{booktabs}
\usepackage{array,ragged2e}
\usepackage[table]{xcolor}
\usepackage{multirow}
\usepackage{algorithm}
\usepackage{algpseudocode}
\usepackage{pifont}

\usepackage{xcolor}

\journal{Expert Systems with Applications}
\begin{document}
\begin{frontmatter}

\title{Grey Models for Short-Term Queue Length Predictions for Adaptive Traffic Signal Control}

\author[MCSaddress]{Gurcan Comert\corref{mycorrespondingauthor}}
\cortext[mycorrespondingauthor]{Corresponding author}
\ead{gurcan.comert@benedict.edu}

\author[PEaddress]{Zadid Khan}
\ead{mdzadik@clemson.edu}

\author[PEaddress]{Mizanur Rahman}
\ead{mdr@clemson.edu}

\author[PEaddress]{Mashrur Chowdhury}
\ead{mac@clemson.edu}

\address[MCSaddress]{Computer Sc., Physics, and Engineering Department, Benedict College, 1600 Harden St., Columbia, SC USA 29204}
\address[PEaddress]{Glenn Department of Civil Engineering, Clemson University, Lowry Hall, Clemson, SC USA 29634}

\begin{abstract}
Traffic congestion at a signalized intersection greatly reduces the travel time reliability in urban areas. Adaptive signal control system (ASCS) is the most advanced traffic signal technology that regulates the signal phasing and timings considering the traffic patterns in real-time in order to reduce traffic congestion. Real-time prediction of traffic queue length can be used to adjust the signal phasing and timings for different traffic movements at a signalized intersection with ASCS. The accuracy of the queue length prediction model varies based on the many factors, such as the stochastic nature of the vehicle arrival rates at an intersection, time of the day, weather and driver characteristics. In addition, accurate queue length prediction for multilane, undersaturated and saturated traffic scenarios at signalized intersections is challenging. Thus,  the objective of this study is to develop short-term queue length prediction models for signalized intersections that can be leveraged by adaptive traffic signal control systems using four variations of Grey systems: (i) the first order single variable Grey model (GM(1,1)); (ii) GM(1,1) with Fourier error corrections (EGM); (iii) the Grey Verhulst model (GVM), and (iv) GVM with Fourier error corrections (EGVM). The efficacy of the Grey models is that they facilitate fast processing; as these models do not require a large amount of data; as would be needed in artificial intelligence models; and they are able to adapt to stochastic changes, unlike statistical models. We have conducted a case study using queue length data from five intersections with adaptive traffic signal control on a calibrated roadway network in Lexington, South Carolina. Grey models were compared with linear, nonlinear time series models, and long short-term memory (LSTM) neural network. Based on our analyses, we found that EGVM reduces the prediction error over closest competing models (i.e., LSTM and Additive Autoregressive (AAR) time series models) in predicting average and maximum queue lengths by $40\%$ and $42\%$, respectively, in terms of Root Mean Squared Error (RMSE), and $51\%$ and $50\%$, respectively, in terms of Mean Absolute Error (MAE).
\end{abstract}


\end{frontmatter}

\section{Introduction}
\label{sctIntro}
Traffic congestion at a signalized intersection negatively impacts the travel time reliability in urban areas (\cite{qi2016impact}, \cite{ma2018developing}). Adaptive signal control systems (ASCS) are the most advanced technology that regulates the phasing as well as red, yellow and green timings considering the traffic patterns (i.e., the arrival rate of vehicles at a signalized intersection from different approaches) in real-time to reduce traffic congestion (\cite{radin2018federal}). Major benefits of ASCS include: (i) real-time distribution of green timings based on the arrival rate of vehicles for all traffic movements; and (ii) reduction of travel times through intersections by ensuring progression through green signal timing window (\cite{radin2018federal}). 

Real-time prediction of traffic queue length can be used to adjust the green timing for different traffic movements. Existing systems mainly use inductive loop detectors to detect queue lengths. Inductive loop detectors are installed on the roadway pavement (\cite{tiaprasert2015queue}). There are several disadvantages of the loop detector-based sensors: (i) low coverage area (only cover a small length of a traffic lane); (ii) detection susceptibility to environmental conditions; and (iii) high cost for deployment and maintenance. Emerging connected vehicle technology can overcome the challenges of existing queue length estimation methods by providing real-time information to the traffic signal control using Vehicle-to-Infrastructure (V2I) wireless communication (\cite{tiaprasert2015queue}).

In a connected vehicle environment, the information of arrival rate of vehicles for all movements at a signalized intersection is available via V2I communication. However, these arrival rates are stochastic in nature depending on different factors, such as the time of the day, weather and driving characteristics (\cite{yang2018impacts}). These factors adversely affect the performance of queue length prediction models and reduce the prediction accuracy significantly. Moreover, accurate queue length predictions for multilane scenarios and robustness of the predictions for both under saturated and saturated roadway traffic scenarios at a signalized intersection are challenging (\cite{zhan2015lane}).

Recent studies use statistical and data-driven models for predicting queue length at signalized intersections (\cite{tiaprasert2015queue}, \cite{comert2016queue}). Data-driven models, such as recurrent neural network (RNN) based time series models, require a large amount of data for training a queue prediction model for different scenarios (such as single lane and multilane roadways) to achieve a high accuracy. However, it increases the computational resource need for real-time applications. It also increases the need for large amounts of data for extensive training considering different roadway  traffic scenarios. The advantage of the RNN models is that after training, it can capture the stochastic  roadway traffic pattern. On the other hand, although statistical models do not require a large amount of data for training, they need to re-estimate model parameters based on the traffic patterns, which reduces the applicability of the statistical model for real-world applications (\cite{comert2016queue}).

Recently, Grey models (GM) have become popular for traffic data prediction, as these models do not assume any underlying distribution for data generation process; are able to handle autocorrelated observations, and require low computational cost (\cite{bezuglov2016short}). Furthermore, GM requires low sample size to update its parameters (as low as only four data points) (\cite{liu2010grey}). A study by An et al. showed that the accuracy of first order single variable Grey Model (GM(1,1)) is higher than back propagation neural network (NN) and radial basis function NN model to predict monthly average daily traffic volume (\cite{an2012exploring}). Similarly, Gao et al. found that GM(1,1) prediction accuracy of average hourly traffic volumes surpasses the performance of support vector machine (SVM) and artificial neural network networks (\cite{gao2010road}). However, there is no study that uses Grey models for predicting traffic queue length using connected vehicle data for ASCS. In addition, the efficacies of the Grey models are that it does not require a large amount of data, and is able to adapt to stochastic changes of the arrival rate of vehicles at a signalized intersection. 

The objective of this study is to develop a robust short-term queue length prediction model for adaptive traffic signal control systems using four variations of Grey Systems: (i) the basic Grey model (GM(1,1)); (ii) GM(1,1) with Fourier error corrections (EGM); (iii) the Grey Verhulst model (GVM), and (iv) GVM with Fourier error corrections (EGVM). Grey models are evaluated using queue length data from five signalized intersections with adaptive traffic signal controls in Lexington, South Carolina. To evaluate the performance of different variation of GMs, we compared GMs to existing linear and nonlinear time series prediction models including long short-term memory (LSTM) model.

The rest of the paper is organized as follows. Section~\ref{sctrw} presents related work focusing on queue length estimations and predictions at signalized intersections. Section~\ref{sctGM} focuses on the Grey models and covers GM(1,1), the Grey Verhulst model, and two variations of these methods to improve their prediction accuracy. Section~\ref{sctexp} presents the compared time series models and detailed numerical experiments to evaluate the prediction performance of the Grey models. Finally, section~\ref{sctconc} summarizes the findings and addresses possible future research directions.
  
\section{Related Work}
\label{sctrw}
There have been many studies focusing on queue length estimations and predictions at signalized intersections. Different studies have used different types of models and inputs for estimating or predicting queue lengths at intersections. Below we segment the literature into prediction and estimation studies.

\subsection{Queue Length Prediction}
\label{scqlp}
In one of the most recent studies, Li et al. developed a queue length prediction model for multi-lane signalized intersections. The authors used the Lighthill-Whitham-Richards shockwave theory and Robertson's platoon dispersion model to predict the arrival of vehicles $5$ seconds in advance for each lane and integrated the predictions of different lanes using Kalman filter. The authors achieved an average RMSE of $2.33$ vehicles, MAE of $1.82$ vehicles, and MAPE of $16.12\%$ for maximum queue length prediction. However, this model does not consider several aspects of real-world traffic flow that affect queue lengths, such as lane changing, heterogeneous traffic and dynamic correction of travel times (\cite{li2018real}). Zeng et al. developed a queue length prediction model using stochastic fluid theory. The authors used the two-fluid theory for considering road traffic and congested traffic for predicting queue lengths. The average relative prediction error of the model is $24.7\%$ for single lane scenario and $38.2\%$ for multilane scenario. This model also struggles with multilane scenario due to the existence of lane changing (\cite{zeng2017research}). 

\subsection{Queue Length Estimation}
\label{scqle}
All these studies have the limitation of only estimating current queue lengths, but these studies are also relevant to prediction because the estimation models can be leveraged to create prediction models. The estimation models can be divided into three categories, statistical, analytical, and data-driven models.

\subsubsection{Statistical models}
\label{scsm} 
Comert developed stochastic models and formulated the analytical expressions of estimators, which were used for estimating queue length from probe vehicle data (e.g., location, time, and count). The developed models estimate cycle-to-cycle queue lengths within $\pm5\%$ error. However, this paper does not deal with predicting future queue lengths (\cite{comert2016queue}). Hao et al. developed seven Bayesian network models for estimating cycle-by-cycle queue lengths for seven different traffic scenarios. The input to the models is mobile traffic sensor data collected between the upstream and downstream of an intersection. Hao et al. proved that the stochastic approach at low penetration rates is more robust compared to deterministic approaches. However, this model suffers from the lack of availability of actual ground truth data, since the model predicts queue length distribution by cycle, but in the real world only a queue is observed at a certain instant (\cite{hao2014cycle}). Zhan et al. developed a lane-based real-time queue length estimation method using license plate data. The developed model includes a Gaussian Process based interpolation method and a car following model for reconstructing the equivalent cumulative arrival-departure curve of each lane and estimating queue lengths. The RMSE and MAE of queue length estimation are below $3.2$ vehicles and $2.4$ vehicles (approximately $12$ m and $16$ m based on average vehicle length), respectively. However, this model also has some limitations, such as lane changing effects not considered and the model may infer incorrect arrival times (\cite{zhan2015lane}).

\subsubsection{Analytical models}
\label{scanm} 
Hao and Ban developed a queue length estimation method to solve the long queue problem using short vehicle trajectories from mobile sensors (\cite{hao2015long}). The method is based on vehicle trajectory reconstruction models to estimate the missing acceleration/deceleration process. Their method was able to reduce the mean absolute error for long queue length estimation from $3.79$ vehicles to $1.61$ vehicles (approximately $18.95$ m to $8.05$ m based on average vehicle length). However, Hao and Ban do not deal with predicting future queue lengths. Moreover, this model is inapplicable for multi-lane intersections and heavily congested scenarios. It also requires the input data to be high precision and low sampling rate (\cite{hao2015long}). Wang et al. developed a queue estimation method for signalized intersections using data from both probe vehicles and point detectors. The authors used shockwave theory to model the queue dynamics. The models showed mean absolute percent error (MAPE) of $17.09\%$ and $12.28\%$ for 2 different scenarios. However, this model has some limitations in estimating queue length when there is residual queue at the intersection (\cite{wang2017shockwave}). Tiaprasert et al.  proposed a queue length estimation model using connected vehicle technology for adaptive signal control (\cite{tiaprasert2015queue}). Tiaprasert et al. applied a discrete wavelet transform (DWT) to queue estimation in order to make it robust against randomness in penetration ratio. The authors showed that the queue length estimation algorithm works in both undersaturated and saturated traffic conditions, which is essential for applying it in adaptive signal control (\cite{tiaprasert2015queue}).

\subsubsection{Data-driven models}
\label{scddm}
An et al. developed a real-time queue length estimation model including a breakpoint misidentification checking process and two input-output models (upstream-based and local-based), and used event-based data as input. The model was able to improve on the generic breakpoint model as the maximum queue length estimation MAE was found to be $10.88$ m compared to $32.2$ m. However, as the model needed to be trained with ground truth data for parameter estimation; two limitations of the model related to parameter estimation are: validity of the parameters for different time periods and transferability of the parameters among intersections (\cite{an2018real}). Gao et al. proposed a cycle-by-cycle queue length estimation model, which is a weighted combination of two submodels: shockwave sensing and back propagation neural network sensing. The input to the model is connected vehicle data. The authors showed that their model has a higher accuracy than probability distribution models for low penetrations of connected vehicles, with $85\%$ accuracy for low penetration rates and $95\%$ accuracy for high penetration rates. This model also performs well for both undersaturated and saturated conditions, which is crucial for adaptive signal control. However, it suffers from the data requirements for training the back propagation neural network model (\cite{gao2019connected}).

From the review of literature, it is evident that queue length prediction has some research gaps, which include accuracy for multilane scenarios and robustness for both under saturated and saturated scenarios (to be effective for adaptive signal control). Through our development and evaluation of Grey model, we will investigate these gaps in the literature.

\section{Grey Systems for Queue Length Prediction}
\label{sctGM}
The Grey Systems theory was developed by \cite{ju1982control} and since then it has become the preferred method to study and model systems in which the structure or operation mechanism is not completely known (\cite{ju1982control}). Grey System theory applications has been applied mainly in the area of finance (\cite{kayacan2010grey}). Its application in transportation is limited; examples include prediction of average speed, travel time, number of accidents, and pavement degradation (\cite{an2018real}, \cite{gao2019connected}, \cite{liu2014highway}, \cite{bezuglov2016short}).
According to the Grey Systems theory, the unknown parameters of the system are represented by discrete or continuous Grey numbers. The theory introduces a number of properties and operations on the Grey numbers, such as the core of the number, its degree of Greyness, and whitenization of the Grey number. The latter operation generally describes the preference of the number towards the range of its possible values (\cite{liu2014highway}).
In order to model time series, the theory suggests a family of Grey models, where the basic one is the first order Grey model with one variable which will be referred to as GM(1,1). The principles and estimation of GM(1,1) is briefly discussed here; readers are referred to Ju-Long (1989) (\cite{julong1989introduction}) for additional information. 

Suppose that $X^{(0)}=(x^{(0)}(1),x^{(0)}(2),...,x^{(0)}(n))$ denotes a sequence of nonnegative observations of a stochastic process (i.e., average and maximum queue lengths) and $X^{(1)}=(x^{(1)}(1),x^{(1)}(2),...,x^{(1)}(n))$ is an accumulation of sequence of queue lengths, $X^{(1)}$ computed as in Eq.~(\ref{eq:accumulated_seq}). If the data contains missing values, sequence of identical observations, or zeros (no queues during parts of green phases), one can introduce very low Gaussian noise $\sim N(0,0.0001)$. 
\begin{equation}
x^{(1)}(k) = \sum_{i=1}^{k}{x^{(0)}(i)}
\label{eq:accumulated_seq}
\end{equation}
then~Eq.~(\ref{eq:original_GM}) defines the original form of the GM(1,1).
\begin{equation}
x^{(0)}(k) + ax^{(1)}(k) = b
\label{eq:original_GM}
\end{equation}
Let $Z^{(1)}=(z^{(1)}(2),z^{(1)}(3),...,z^{(1)}(n))$ be a mean sequence of $X^{(1)}$ calculated by formula Eq.~(\ref{eq:Z_1}) and defined for $k = 2,3,\cdots,n$
\begin{equation}
z^{(1)}(k) = \frac{z^{(1)}(k-1)+z^{(1)}(k)}{2}
\label{eq:Z_1}
\end{equation}
Eq.~(\ref{eq:basic_GM}) gives the basic form of GM(1,1).
\begin{equation}
x^{(0)}(k) + az^{(1)}(k) = b
\label{eq:basic_GM}
\end{equation}
If $\hat{a}=(a,b)^T$ and 
\begin{eqnarray*}
Y = \left[ 
\begin{array}{c}
x^{(0)}(2) \\
x^{(0)}(3) \\
\vdots \\
x^{(0)}(n) \\
\end{array} 
\right],
B = \left[ 
\begin{array}{cc}
-z^{(1)}(2) & 1 \\
-z^{(1)}(3) & 1 \\
\vdots & \vdots \\
-z^{(1)}(n) & 1 \\
\end{array} 
\right]. 
\label{eq:YandB}
\end{eqnarray*}
then, as in~\cite{liu2006grey}, the least squares estimate of the GM(1,1) model is~$\hat{a}=(B^TB)^{-1}B^TY$ and Eq.~(\ref{eq:whitenization_GM}) is the whitenization equation of the GM(1,1) model.
\begin{equation}
\frac{dx^{(1)}}{dt} + ax^{(1)}(k) = b
\label{eq:whitenization_GM}
\end{equation}

Suppose that $\hat{x}^{(0)}(k)$ and $\hat{x}^{(1)}(k)$ represent the time response sequence (the forecast) and the accumulated time response sequence of GM(1,1) at time $k$, respectively. Then, the accumulated time response sequence can be obtained by solving Eq.~(\ref{eq:whitenization_GM}): 
\begin{equation}
\hat{x}^{(1)}(k+1)=\left(x^{(0)}(1)-\frac{b}{a}\right)e^{-ak}+\frac{b}{a}, k=1,2,...,n
\label{eq:model_x_1_solution}
\end{equation}
According to the definition in Eq.~(\ref{eq:accumulated_seq}), the restored values of $\hat{x}^{(0)}(k+1)$ are calculated as $\hat{x}^{(1)}(k+1)-\hat{x}^{(1)}(k)$:
\begin{equation}
\hat{x}^{(0)}(k+1)=\left(1-e^a\right)\left(x^{(0)}(1)-\frac{b}{a}\right)e^{-ak}, k=1,2,...,n
\label{eq:model_x_0_solution}
\end{equation}
Eq.~(\ref{eq:model_x_0_solution}) gives the method to produce forecasts for all $k$ in $2,3,...,n$. However, for longer time series, a rolling GM(1,1) is preferred. The rolling model observes a window of a few sequential data points in the series: $x^{(0)}(k+1),x^{(0)}(k+2),...,x^{(0)}(k+w)$, where $w \geq 4$ is the window size. Then, the model forecasts one or more future data points: $\hat{x}^{(0)}(k+w+1), \hat{x}^{(0)}(k+w+2)$. The process repeats for the next $k$ value. 

\subsection{The Grey Verhulst Model (GVM)}
\label{sctGVM}
The response sequence Eq.~(\ref{eq:model_x_0_solution}) implies that the basic GM(1,1) works the best when the time series demonstrate a steady growth or decline and may not perform well when the data has oscillations or saturated sigmoid sequences. For the latter case, the Grey Verhulst model (GVM) is generally used~(\cite{liu2010grey}). The basic form of the GVM is present by Eq.~(\ref{eq:verhulst_model}).
%
%
\begin{equation}
x^{(0)}(k)+az^{(1)}(k)=b\left(z^{(1)}(k)\right)^2
\label{eq:verhulst_model}
\end{equation}
Eq.~(\ref{eq:verhulst_whitenization}) provides the whitenization formula of GVM. It is practically represents assumed structure of data generation process.
\begin{equation}
\frac{dx^{(1)}}{dt} + ax^{(1)} = b\left(x^{(1)}\right)^2
\label{eq:verhulst_whitenization}
\end{equation}

Similar to the GM(1,1), the least squares estimate is applied to find~$\hat{a}=(B^TB)^{-1}B^TY$, where 

\begin{eqnarray*}
Y = \left[ 
\begin{array}{c}
x^{(0)}(2) \\
x^{(0)}(3) \\
\vdots \\
x^{(0)}(n) \\
\end{array} 
\right],
B = \left[ 
\begin{array}{cc}
-z^{(1)}(2) & z^{(1)}(2)^2 \\
-z^{(1)}(3) & z^{(1)}(3)^2 \\
\vdots & \vdots \\
-z^{(1)}(n) & z^{(1)}(n)^2 \\
\end{array} 
\right]. 
\label{eq:YandB_VerhulstModel}
\end{eqnarray*}

The forecasts $\hat{x}^{(0)}(k+1)$ are calculated using Eq.~(\ref{eq:verhulst_x_0_solution}).

\begin{equation}
\hat{x}^{(0)}(k+1)=\frac{ax^{(0)}(1)\left(a-bx^{(0)}(1)\right)}{bx^{(0)}(1)+\left(a-bx^{(0)}(1)\right)e^{a(k-1)}}*\frac{\left(1-e^a\right)e^{a\left(k-2\right)}}{bx^{(0)}(1)+\left(a-bx^{(0)}(1)\right)e^{a(k-2)}}
\label{eq:verhulst_x_0_solution}
\end{equation}
%

\subsection{Error Corrections to Grey Models}
\label{scterror}
The accuracy of the Grey models can be improved by a few methods. Suppose that $\epsilon^{(0)}$=$\epsilon^{(0)}(1),...,\epsilon^{(0)}(n)$ is the error sequence of $X^{(0)}$, where $\epsilon^{(0)}(k)$= $x^{(0)}(k)-\hat{x}^{(0)}(k)$. If all errors are positive, then a remnant GM(1,1) model can be built (\cite{liu2010grey}). When the errors can be positive or negative, $\epsilon^{(0)}$ can be expressed using Fourier series (\cite{tan1996residual}) as in Eq.~(\ref{eq:fourier_series}).
%
%
%
%
%
\begin{equation}
\epsilon^{(0)}(k) \cong \frac{1}{2}a_0+\sum_{i=1}^{z}\left[a_i cos\left(\frac{2\pi i}{T}k\right)+b_i sin\left(\frac{2\pi i}{T}k\right) \right], k = 2,3,...,n
\label{eq:fourier_series}
\end{equation}
where $T=n-1$ and $z=\left( \frac{n-1}{2}\right)-1$.

The solution is found via the least squares estimate, presuming that $\epsilon^{(0)} \cong PC$ where C is a vector of coefficients: $C=\left[a_0, a_1 b_1, a_2 b_2... a_n b_n \right]^T$ and matrix P is:
\begin{eqnarray*}
P = \left[ 
\begin{array}{c c c c c c}
\frac{1}{2} & cos\left(2\frac{2\pi}{T}\right) & sin\left(2\frac{2\pi}{T}\right) & \dots & cos\left(2\frac{2\pi z}{T}\right) & sin\left(2\frac{2\pi z}{T}\right)\\
\frac{1}{2} & cos\left(3\frac{2\pi}{T}\right) & sin\left(3\frac{2\pi}{T}\right) & \dots & cos\left(3\frac{2\pi z}{T}\right) & sin\left(3\frac{2\pi z}{T}\right)\\
\vdots & \vdots & \vdots & \vdots & \vdots & \vdots\\
\frac{1}{2} & cos\left(n\frac{2\pi}{T}\right) & sin\left(n\frac{2\pi}{T}\right) & \dots & cos\left(n\frac{2\pi z}{T}\right) & sin\left(n\frac{2\pi z}{T}\right)\\
\end{array} 
\right] 
\label{eq:Fourier_P}
\end{eqnarray*}
%
%
then $C \cong \left(P^T P \right)^{-1}P^T\epsilon^{(0)}$. As a result, the predicted value of the time series must be corrected according to Eq.~(\ref{eq:fourier_correction}):
\begin{equation}
\hat{x}_f^{(0)}(k) = \hat{x}^{(0)}(k) + \epsilon^{(0)}(k), k = 2,3,...,n
\label{eq:fourier_correction}
\end{equation}
\section{Numerical Experiments}
\label{sctexp}
In this section, numerical results are presented regarding the performance of the applied Grey models (GMs). This section also presents linear and nonlinear time series models and their parameter values to compare the prediction results with Grey models. 
\subsection{Data Description}
\label{sctDD}
\begin{figure}[h!]
\centering
\includegraphics[scale=.4]{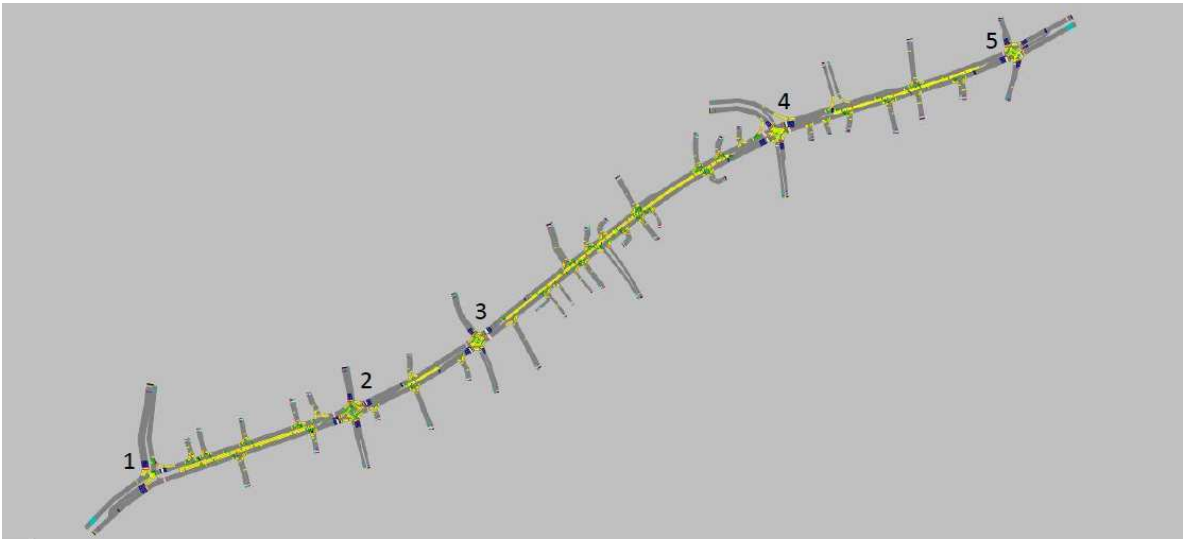}
\caption{US 378 Lexington, South Carolina corridor with intersections (1 to 5) simulated in VISSIM}
\label{fig_1}       
\end{figure}

In order to evaluate the performance of the Grey System models, a case study has been performed. A calibrated microsimulation model has been developed in VISSIM of the US 378 (Sunset Drive) corridor in Lexington, South Carolina. A portion of the corridor has been chosen for analysis that includes five signalized intersections. All the signalized intersections operate under adaptive signal control. Centracs Adaptive traffic signal controller has been used in this study. Centracs Adaptive is an improved version of the original ACS Lite controller developed by Econolite. Traffic data and travel times have been collected for afternoon peak period and the VISSIM model has been calibrated to this data. As we are interested in queue lengths, a congested scenario is required in order to study the patterns of queue buildups and progressions. The first intersection is a T-intersection, while the other four are 4-way intersections. Along with the five intersections, there are 33 driveways on this corridor, which creates disruptions and stop-go conditions. These can contribute to the queue length patterns at intersections. A screenshot of the VISSIM simulation environment is shown in Fig.~\ref{fig_1}, including the detectors and queue counters placed at intersections.
\begin{table}[!htb]
\centering
\caption{Queue Counter Information}
\label{tab_dataset}
\scalebox{0.6}{
\begin{tabular}{l|c c c c c c c c c c c c c c c c c c c c c c c c c c c c c c c}
\hline\noalign{\smallskip}
 Queue Counter & 1&2 &3 &4 &5 &6 &7 &8 &9 &10 &11 &12 &13 &14 &15 &16 &17 & 18 & 19 &20 &21 &22 &23 &24 &25 &26 &27 &28 &29 &30 &31 \\
\noalign{\smallskip}\hline\noalign{\smallskip}
Number of Lanes & 2&2 &2 &3 &2 &1 &1 &2 &1 &2 &2 &2 &1 &1 &2 &1 &2 & 2 & 1 &1 &1 &2 &1 &2 &2 &1 &1 &1 &2 &1 &1 \\
\noalign{\smallskip}\hline\noalign{\smallskip}
Intersection &  1&1 &1 &1 &2 &2 &2 &2 &2 &2 &3 &3 &3 &3 &3 &3 &4 & 4 & 4 &4 &4 &4 &4 &5 &5 &5 &5 &5 &5 &5 &5\\
\noalign{\smallskip}\hline\noalign{\smallskip}
\end{tabular}
}
\end{table}

In order to get the queue length data, queue counters are placed at each intersection. Each queue counter corresponds to one lane group. A lane group is a group of lanes that allow traffic to move simultaneously. For example, a through lane and a right-turn lane can be in the same lane group. However, a through lane and a left-turn lane may not be in the same lane group, as the two lanes may not allow traffic to flow simultaneously. There are 31 queue counters in total for five intersections. By running the simulation, we have collected the average and maximum queue length data for each queue counter at each intersection. Please note that some queue counters correspond to multiple lanes. That is why, we have divided the dataset into two segments: average queue length data and maximum queue length data. The intersections are numbered from west to east. Intersection 1 is a T-intersection, so it requires the least number of queue counters (i.e. 4). The queue length data has been collected per second. The information about queue counters is given in Table~\ref{tab_dataset}.
\begin{figure}[h!]
\centering
  \includegraphics[width=0.75\linewidth]{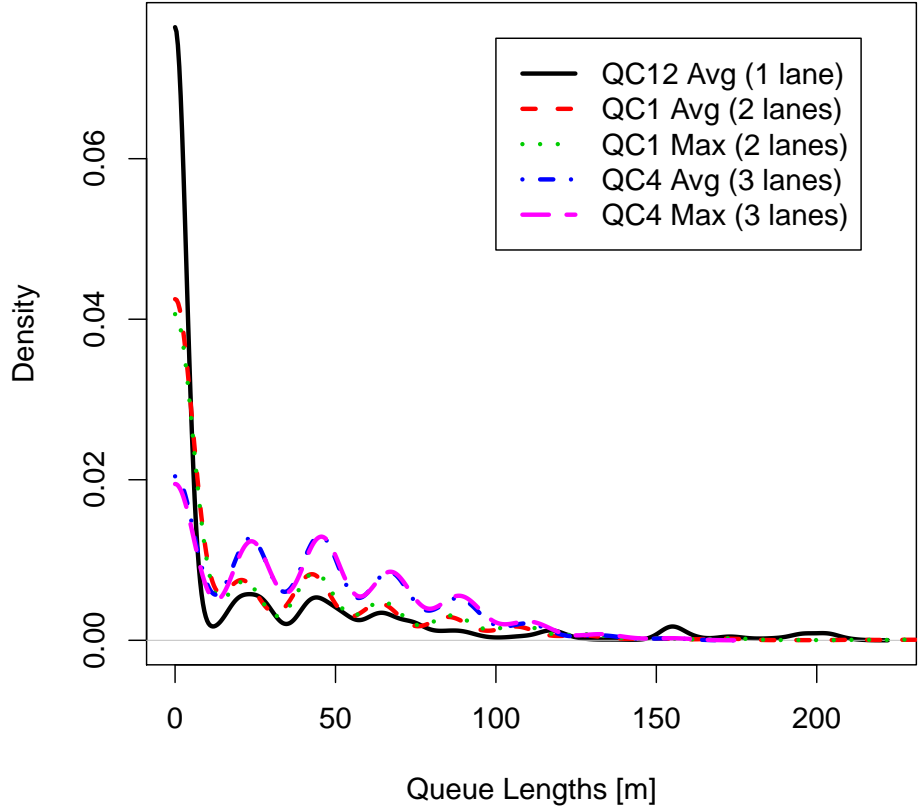}
  \caption{Comparison of average (Avg) and maximum (Max) queue length densities for different Queue Counters (QC1,QC4, and QC12)}
\label{fig_2}
\end{figure}

Average and maximum queue length data of all 31 counters is collected for 1 hour from 4 different simulation runs. From Table~\ref{tab_dataset}, it can be observed that different queue counters yield different types of queue length patterns based on number of lanes, intersection, signal phasing and timings etc. For example, in the case of queue counter number 6 (denoted as QC6), the number of lanes is 1, so the average and maximum queue length is the same. However, for QC4, the number of lanes is 3, so there will be variations between the maximum and average queue lengths. The difference between average and maximum queue length of QC4 (3 lanes) is shown in Fig.~\ref{fig_2}) and the variation of average queue lengths among 3 different queue counters, QC12 (1 lane), QC1 (2 lanes) and QC3 (3 lanes), is shown in Fig.~\ref{fig_2}. From Fig.~\ref{fig_2}, it can be observed that the variation in average queue length is higher than the maximum queue length, which indicates the existence of one more congested lane compared to the other lanes in the lane group. From Fig.~\ref{fig_2}, it can be observed that the queue buildup for QC12 is more severe at certain times, which takes time to dissipate. On the other hand, QC1 and QC4 have a more distributed queue accumulation and dissipation due to the higher number of lanes.

\begin{figure}[h!]
\centering
\includegraphics[scale=.35]{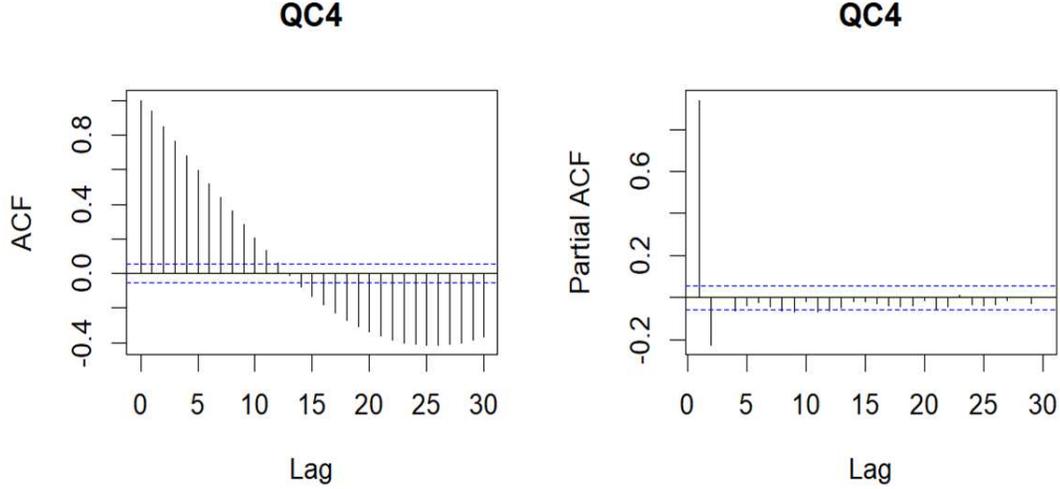}
\caption{Example of Autocorrelation functions (ACF) of queue length time series data on Queue Counter 4 (QC4)}
\label{fig_3}       
\end{figure}

Autocorrelation presence within the time series data assists in prediction if the models can capture them. Although several other covariates would influence (hidden or unobserved) the response variable of interest, we can simply use historical data to be able to predict future values. These conditions constitute the main motivation behind Grey system models. The autocorrelations can be shown simply using autocorrelation functions (ACF) and partial ACF or formal statistical tests. As an example, for QC4 average queue lengths, Fig.~\ref{fig_3} shows the presence of negative autocorrelation in the data (Fig.~\ref{fig_3}). The partial ACF plot also reveals that ACF values become insignifiant after 2 significant lags which suggests that the autoregressive (AR) component in the time series to be fit is low (e.g.,AR(1) to AR(3)). A formal Durbin-Watson test also results in a p-value of $0.056$, which barely rejects the null hypothesis of no autocorrelation. Although other parts of data used may fail to reject, the queue length data from our experiments show autocorrelations. 
\subsection{Linear and Non-linear Methods for Comparison}
\label{sctCFM}
Based on the above discussions, time series models can be good forecasting model candidates. For a fair comparison, we considered the following linear and nonlinear time series models for comparison with the GM models which are adopted from (\cite{di2015package}), and long short term memory (LSTM) model. All these models are fit to 1st $67\%$ of all queue counter datasets and tested on $33\%$ of the data that contains $124$ $1$-hour series.

Eq.(\ref{eqn_linear}) presents linear model (autoregressive AR(3) denoted as LINEAR).
\begin{equation}
Z_{t+1}=\mu+\phi_1 Z_t+\phi_2 Z_{t-1}+\phi_3 Z_{t-2}+a_{t+1}
\label{eqn_linear}
\end{equation}

where $Z_t$ denotes average or maximum queue length observations at time $t$, $\mu$ is intercept, $\phi_1$, $\phi_2$, and $\phi_3$ are weights of previous observations, and $a_{t+1}$ is white noise.

Eq.~(\ref{eqn_lstar}) shows logistic smooth transition autoregressive model (LSTAR).
\begin{eqnarray}
Z_{t+1}=\begin{cases}(\phi_1+\phi_{10}Z_t+\phi_{11}Z_{t-\delta}+...+\phi_{1L}Z_{t-(L-1)\delta})(1-G(Z_t,\gamma,th))+ \\
(\phi_2+\phi_{20}Z_t+\phi_{21}Z_{t-\delta}+...+\phi_{2H}Z_{t-(H-1)\delta})G(Z_t,\gamma,th)+\epsilon_{t+1}\end{cases}
\label{eqn_lstar}
\end{eqnarray}

where $Z_t$ denotes average or maximum queue length at time $t$ and $G(Z_t,\gamma,th)=[1+e^{-\gamma(Z_{t-1}-th)}]^{-1}$ is logistics transition function. $L$ = $1$ to $5$ and $H$ = $1$ to $5$ are low and high regimes, $\delta$ is delay of the transition variable, and $th$ is the threshold value.

Eq.(\ref{eqn_nnts}) gives neural network nonlinear autoregressive model (NNETS). 
\begin{equation}
Z_{t+1}=\beta_0+\sum_{j=1}^D{\beta_ig(\gamma_{0j}+\sum_{i=1}^m{\gamma_{ij}(Z_{t-(i-1)\delta})})}
\label{eqn_nnts}
\end{equation}

where $m$ denotes embedding dimension, $D$ is number of hidden layers of the neural network, and $\beta_i$,$\gamma_{0j}$,$\gamma_{ij}$ represent the weights.

Eq.(\ref{eqn_aar}) presents additive nonlinear autoregressive model (AAR).
\begin{equation}
Z_{t+1}=\mu+\sum_{j=0}^{m-1}{s_j(Z_{t-(j)\delta})}
\label{eqn_aar}
\end{equation}

where $s$ represents nonparametric univariate smoothing functions that depends on $Z_t$s and $\delta$ is the delay parameter. Splines from Gaussian family are fitted in the form of $Z_{t+1}\sim\sum_{i=0}^{m-1}{s(Z_{t},..,Z_{t-j})}$. 

Different number of layers (i.e., $m$ values), which categorize models, are fitted and based on their Akaike information criteria (AIC) values, the best models are selected at each run. Similar analysis and justification can be found in Bezuglov and Comert (\cite{bezuglov2016short}). The models that have been used for comparison are: (1) LINEAR (AR) $m = 3$, (2) LSTAR $m = 3$, (3) NNETS $D = 4$, $m = 4$ ($25$ batch size), (4) AAR model $m = 3$ and (5) LSTM model (200 epochs, 5 LSTM neurons, ReLu activation and 2 lag steps). An arbitrary example of resulting fitted models on queue length data are presented in Table~\ref{tab_nlfits}. 

\begin{table}[h!]
\centering
\caption{Parameters for nonlinear models trained on QC31 max queue lengths}
\label{tab_nlfits}       
\scalebox{0.7}{
\begin{tabular}{l l | c}
\hline\noalign{\smallskip}
& Model & Parameters \\
\noalign{\smallskip}\hline\noalign{\smallskip}
\multirow{3}{*}{LSTAR(2,2,2)-Loop}& L & $\phi_1=0.379$  $\phi_{10}=0.948$  $\phi_{11}=0.198$  $\phi_{12}=-0.015$ \\
& H &  $\phi_2=2.019$  $\phi_{20}=-0.001$  $\phi_{21}=-0.311$  $\phi_{22}=-0.010$ \\
& th & $X_t$=$Z_{t}$, th= 17.74, g=100\\ \hline
\multirow{1}{*}{LINEAR(AR(3))}&  & $\mu=0.228$ $\phi_1=0.979$  $\phi_{2}=0.016$  $\phi_{3}=-0.039$  \\  \hline
\end{tabular}
}
\end{table}
\subsection{Results}
\label{sctrd}
This section describes the findings related to queue length predictions using different models and comparison with the Grey models. The results of our analyses are presented in the following subsections. 
\subsubsection{Overall comparison}
\begin{figure}[h!]
\centering
\includegraphics[scale=0.85]{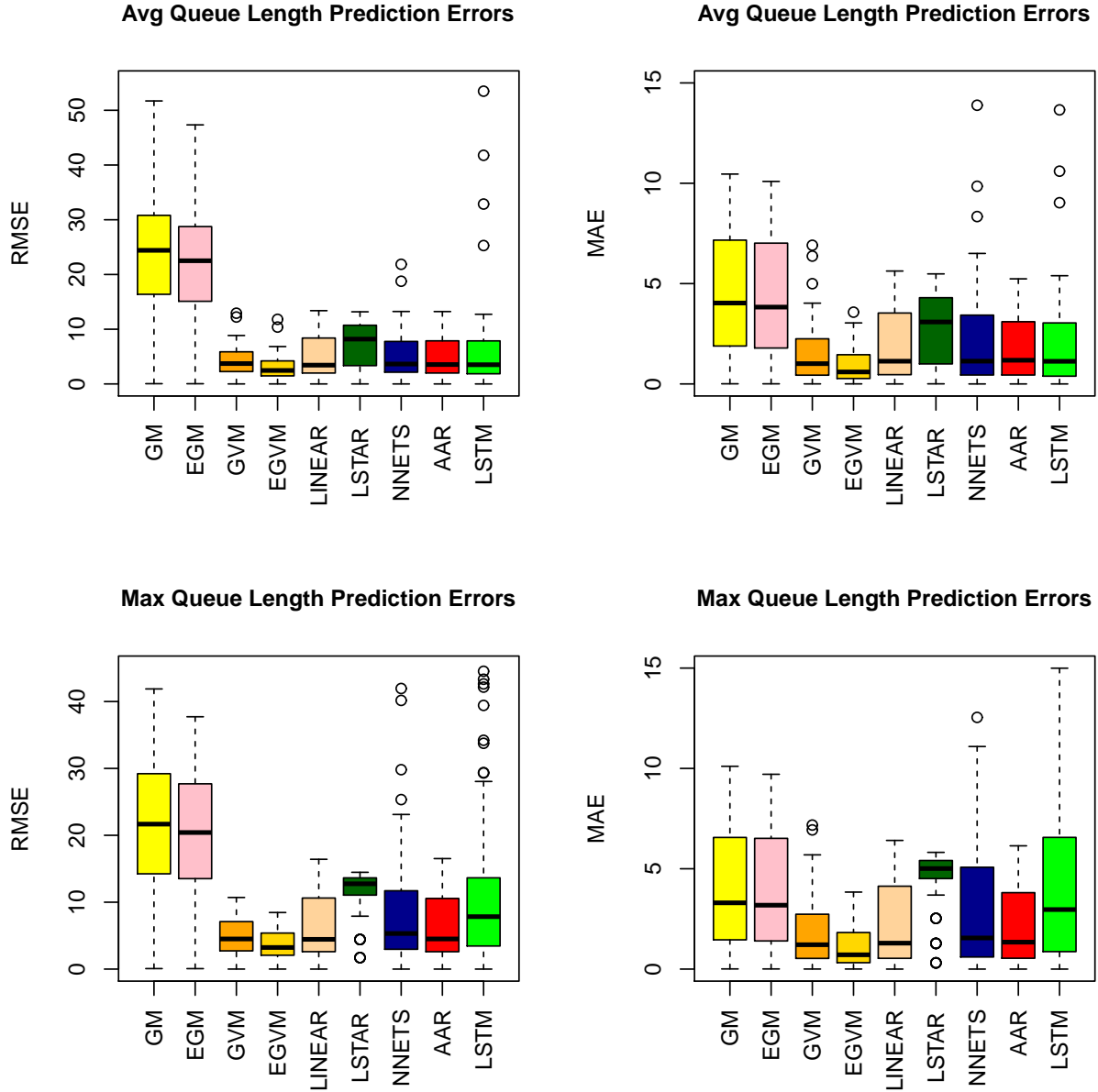}
\caption{Comparison between the performance of different models in predicting queue length}
\label{fig_4}       
\end{figure}

Fig.~\ref{fig_4} demonstrates average and maximum queue length prediction errors in terms of RMSE$=\sqrt{\frac{\sum_{k=1}^{N}{\left(\hat{x}_k-x_k\right)^2}}{N}}$ and MAE$=\frac{1}{N}\sum_{k=1}^{N}{\left|\hat{x}_k-x_k\right|}$. Fig.~\ref{fig_4} contains box plots of RMSE and MAE for all models. Simple GM model and GM model with error corrections do not perform well in predicting queue length data. This is mostly due to the traffic signal generating periodic data. GM model performance can improve if queue lengths are predicted cycle-by-cycle without considering zero queue length. GVM and EGVM are able to capture periodicity with quadratic structures, thus, predicting with higher accuracy compared to GM and EGM models. As shown in Fig.~\ref{fig_4}, LSTM, LINEAR, LSTAR, NNETS, and AAR show similar performance. Their prediction accuracy is much higher than GM and EGM models, but lower than GVM and EGVM models. All models show slightly worse accuracy for maximum queue length prediction compared to average queue length prediction because the variability of the maximum queue lengths have more randomness. 

Fig.~\ref{fig_4} exhibits the following average errors: Avg QLRMSE = [22.83, 21.25, 4.10, 2.94, 5.01, 6.78, 5.08, 4.96, 5.71], Max QLRMSE=[21.09, 19.98, 4.86, 3.69, 6.40, 11.73, 7.77, 6.47, 13.10], Avg QLMAE=[4.43, 4.30, 1.42, 0.91, 1.95, 2.64, 2.06, 1.85, 2.44], and Max QLMAE= [3.96, 3.90, 1.71, 1.10, 2.22, 4.56, 3.23, 2.22, 7.49]). Results show that GVM and EGVM models are able to achieve an error-bound of $\pm5$m in terms of RMSE and $\pm1$ m in terms of MAE for both average and maximum queue lengths. Compared models are able to achieve $\pm2$ m error bound in terms of MAE for average queue length. However, for maximum queue lengths, the error bound increases to $\pm8$ m in terms of MAE. Therefore, GVM and EGVM models are more accurate and robust across all scenarios and error types.    

\begin{figure}[h!]
\centering
\includegraphics[scale=5.0]{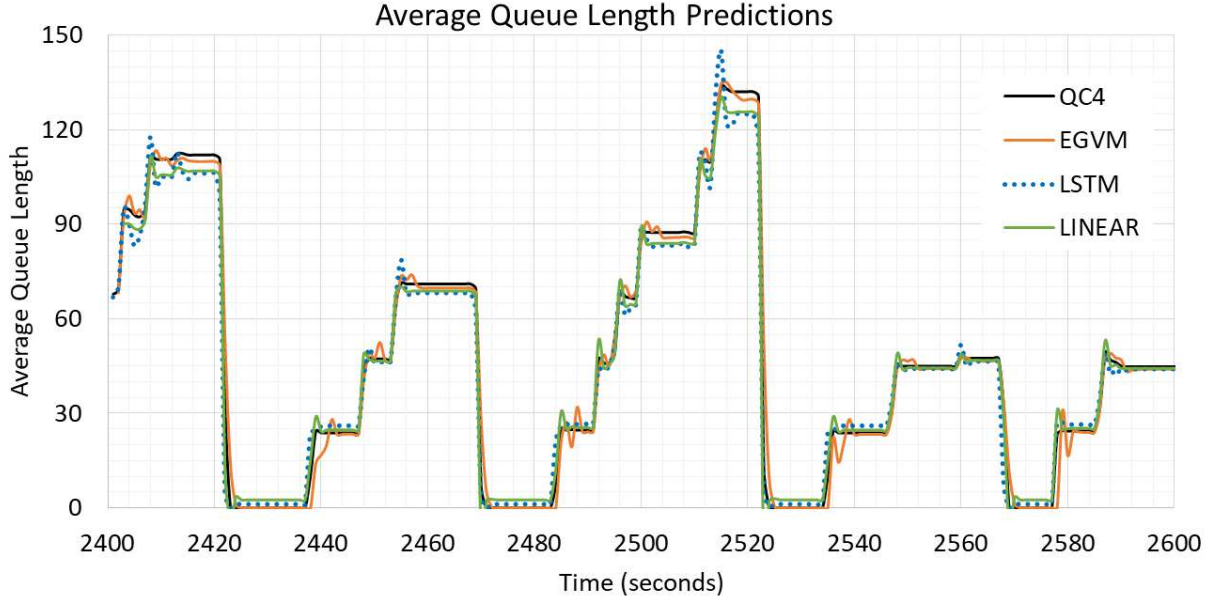}
\caption{Comparison between the performance of EGVM, LSTM, and LINEAR models in predicting average queue lengths of QC4 for 1-step predictions}
\label{fig_5}       
\end{figure}

\begin{table}[htb]
\centering
\small
\caption{Average computational times (in seconds) of different models}
\label{tab_comp}
\scalebox{0.95}{
\begin{tabular}{l l l |l c c c |c c c c | c}
\hline\noalign{\smallskip}
& &Data and Error & GM & EGM & GVM& EGVM & LINEAR & LSTAR & NNETS & AAR & LSTM \\
\noalign{\smallskip}\hline\noalign{\smallskip}
\multirow{2}{*}{Train} & & Avg&- & - & - & - & 0.016 & 7.160 & 0.513 & 0.042 & 50.066 \\
& & Max & -	& - & - & - &	0.016	& 6.580	& 0.421 & 0.044 & 49.883\\

\multirow{2}{*}{Test} & & Avg & 0.115	& 0.863	& 0.128	& 0.883	& 0.391	& 0.499	& 0.480 & 4.696 & 0.958 \\
& & Max & 0.109&	0.812 &	0.115&	0.848	& 0.393 &	0.482	& 0.473 & 4.583 & 0.958 \\
\noalign{\smallskip}\hline\noalign{\smallskip}
\end{tabular}
}
\end{table}
Fig.~\ref{fig_5} presents the comparison between the performance of EGVM, LSTM, and LINEAR models in predicting average queue lengths of QC4 as an example. We observe that LSTM overestimates when there are any abrupt changes in queue lengths. On the other hand, EGVM is able to capture sudden changes in queue length. LINEAR model shows almost similar behavior to LSTM. The reason is that the LSTM model that we have used in this study is a basic model with minimum features (univariate single-step prediction). Moreover, LSTM is a data intensive model but limited ($1$ hour) data has been included in our study.  
 
Lastly, computational times are provided in Table~\ref{tab_comp} per $3600$ observations across all data, training time using $2400$ and testing time of $1200$ observations. GM models do not require any training time and they are updated with low window size (of 4 past observations). LSTM requires more time to learn from the data. Clearly, EGVM is the best option considering both accuracy and computational time. For robust, adaptive, and accurate prediction with low computational times and low sample size, GVM and EGVM models provide accurate prediction of queue length.  

\subsubsection{Model performance comparison (single lane vs multilane)}
\begin{figure}[h!]
\centering
  \includegraphics[width=0.95\linewidth]{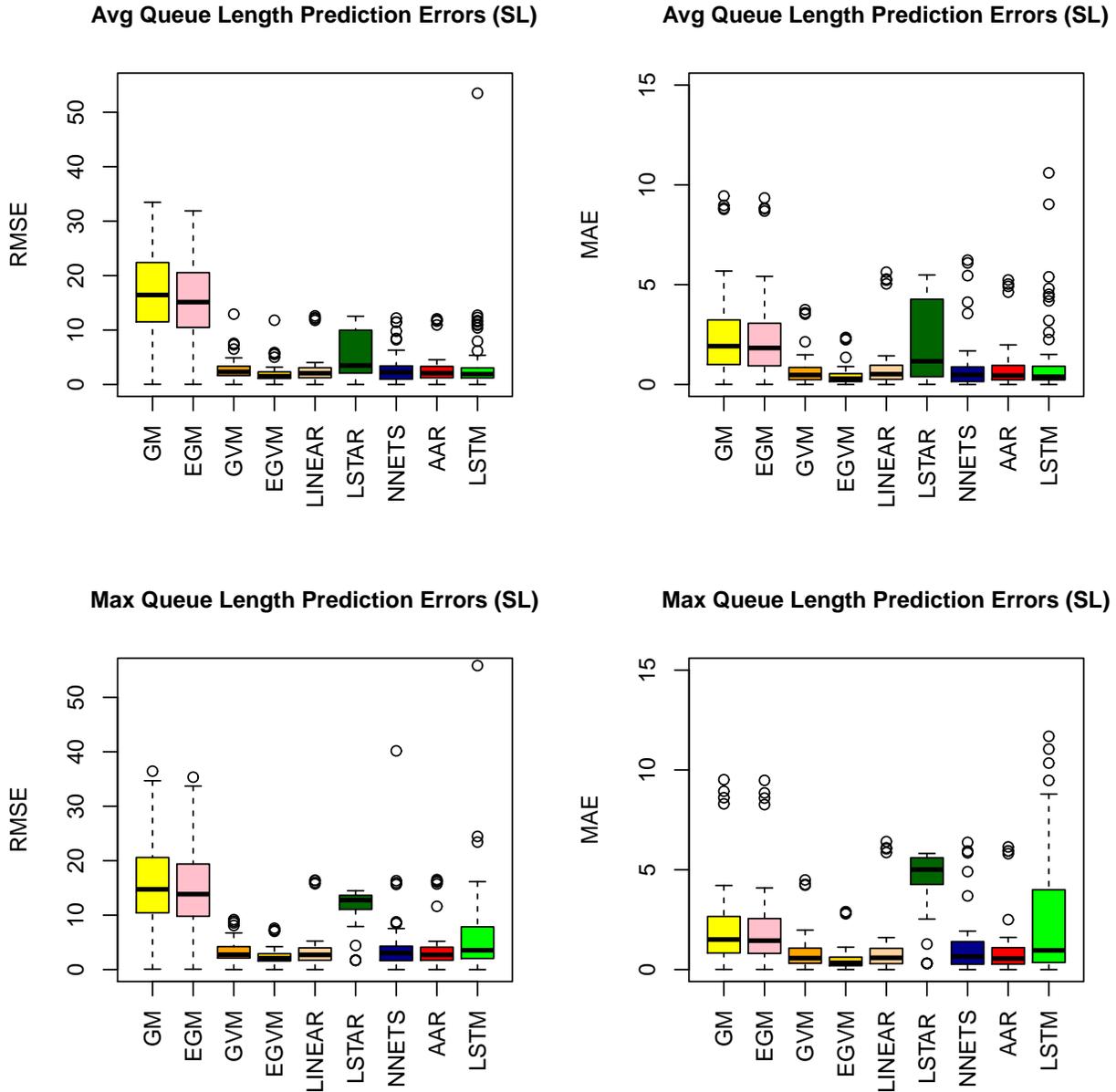}
\caption{Prediction performances on single lane scenarios}
\label{fig_6a} 
\end{figure}

\begin{figure}[h!]
 \centering
\includegraphics[width=0.95\linewidth]{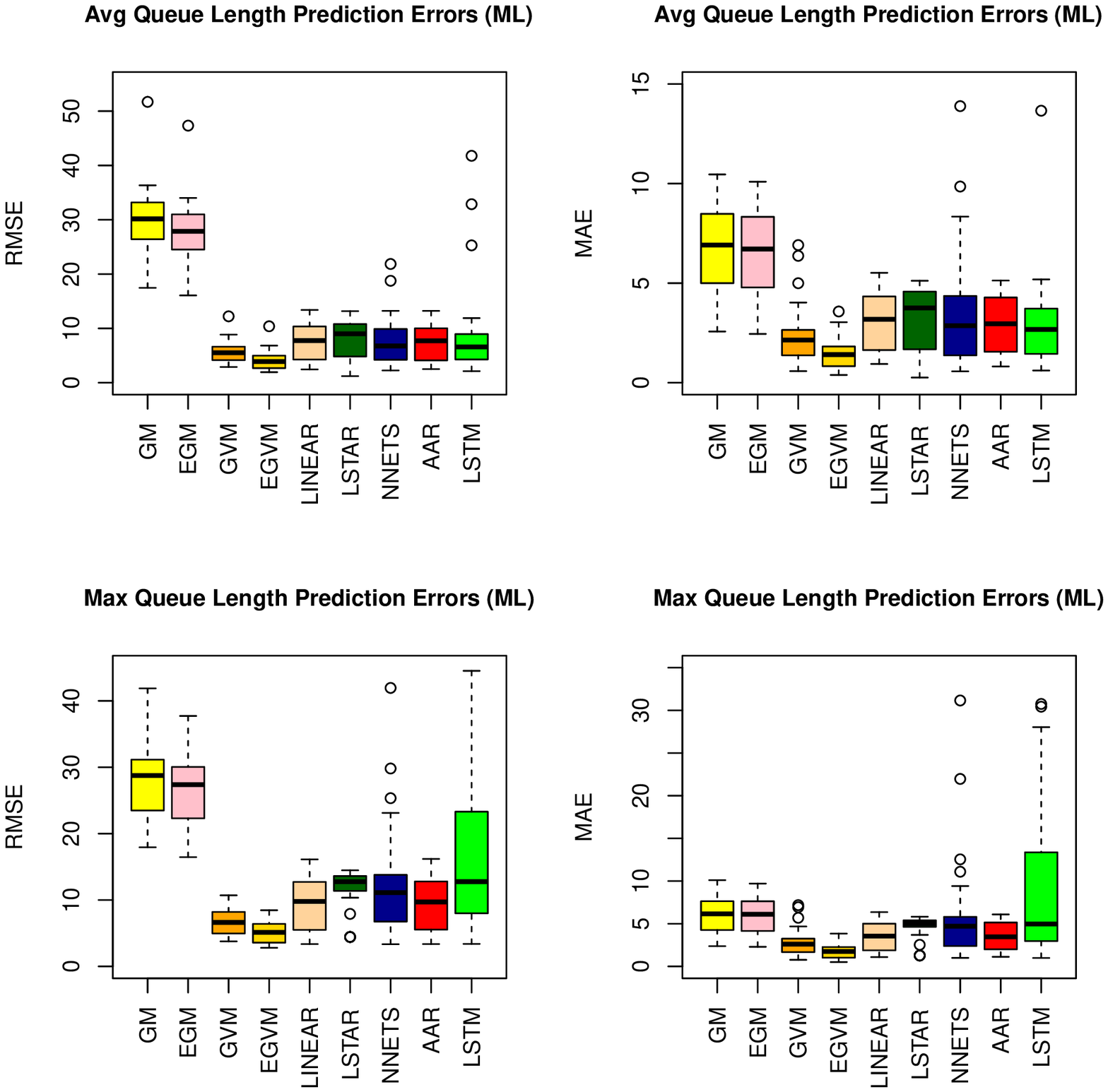}
  \caption{Prediction performances on multi-lane scenarios}
\label{fig_6b}
\end{figure}

As stated in the literature review, we found that two major challenges of the queue length prediction models are their prediction capability for multilane scenarios compared to single lane scenarios and their performance in undersaturated and saturated scenarios. Fig.~\ref{fig_6a} shows all queue length prediction results for single lane scenarios and Fig.~\ref{fig_6b} shows all queue length prediction results for multilane scenarios. Overall, all model performances degrade in multilane scenarios due to many factors; e.g., lane changing behavior of arriving vehicles. However, the EGVM model is still able to maintain a reasonable accuracy for multilane scenarios compared to single lane scenarios. The average RMSE of EGVM model for multilane scenario is $3.55$ m compared to $1.88$ m for single lane scenarios. The average MAE of EGVM model for multilane scenarios is $1.10$ m, compared to $0.44$ m for single lane scenarios. These errors indicate that the EGVM model can be used for both single lane and multilane scenarios.

\subsubsection{Model performance comparison (undersaturated vs saturated)}
\begin{figure}[h!]
\centering
\includegraphics[scale=.3]{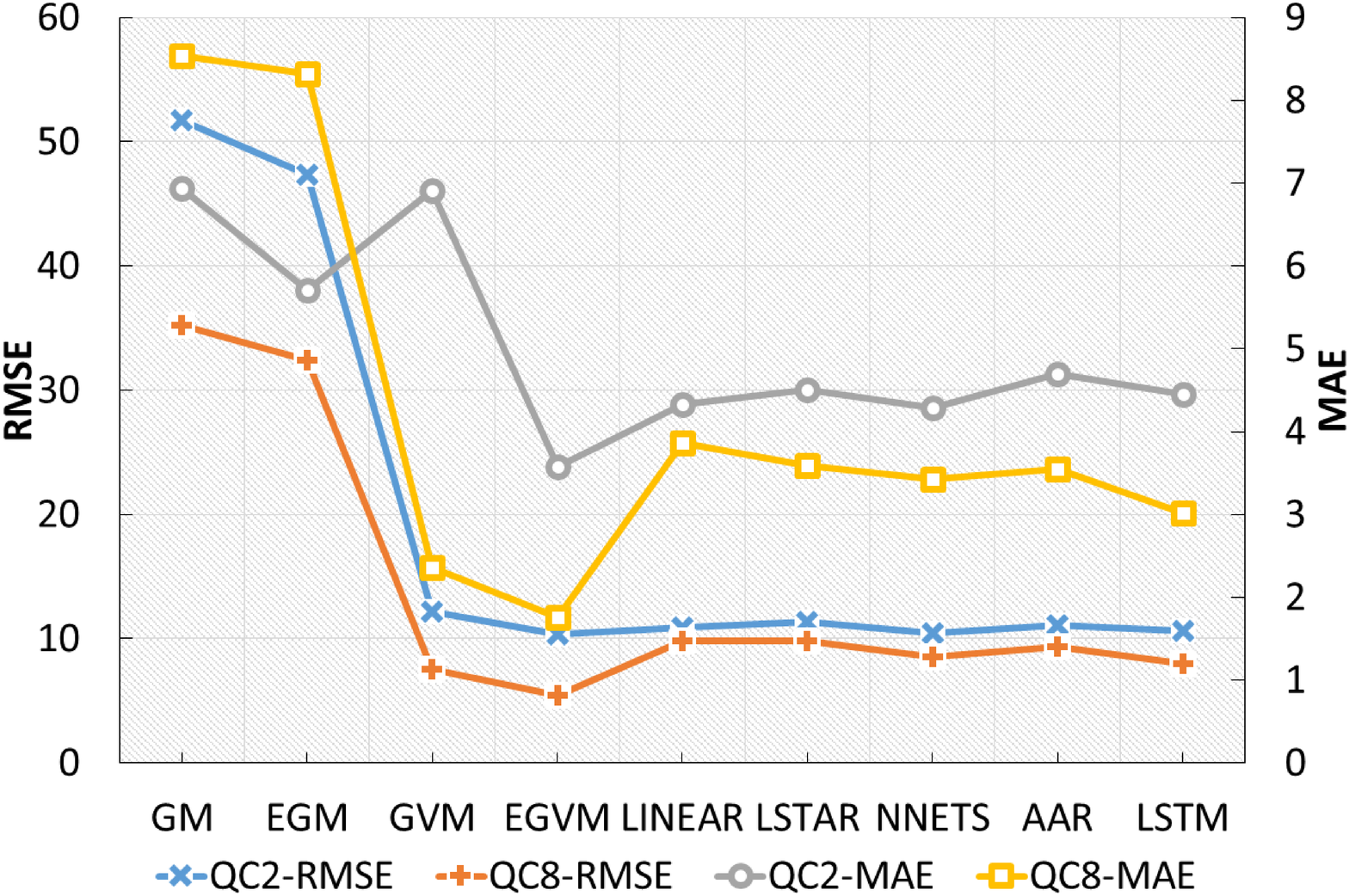}
\caption{Performance of Grey System and other models for undersaturated and saturated scenarios}
\label{fig_7}       
\end{figure}

Within multilane scenarios, we also investigated one queue counter that is operating in saturated conditions, QC2, and another queue counter that is operating in undersaturated conditions, QC8. The comparison of RMSE and MAE values is shown in Fig.~\ref{fig_7}. From Fig.~\ref{fig_7}, we observed that the EVGM model has shown similar performance compared to other models for QC2. However, it has significantly better performance compared to other models for QC8. The RMSE and MAE for QC8 are lower than QC2, which is expected as the congested scenario will create operational issues, such as residual queue and spillback, which could decrease the accuracy of the model. Therefore, the EGVM model can predict queue length with high accuracy in undersaturated conditions while maintaining accuracy comparable to other models for saturated (or congested) conditions. 

\section{Conclusions}
\label{sctconc}
This study shows the effectiveness of Grey Systems in queue length prediction. The EGVM model provides the most accurate queue length predictions for different traffic conditions in both single lane and multilane scenarios. The EGVM model can predict accurately for both undersaturated and saturated conditions, which establish the efficacy of the model for predicting queue length and using it as an input to the adaptive signal control systems. The EGVM model is identified as the best model because it outperforms the compared models for average and maximum queue length prediction. The analysis showed that GVM models could provide approximately 1 meter precision in queue length prediction. Both GVM models provide more accurate prediction than LSTM using only a fraction of the input data (4 vs 2400 observations) and require very low computational times due to the absence of the training phase. This study also showed that simple GM(1,1), even with error correction, failed to produce competitive results compared to other linear and nonlinear time series prediction models. One limitation of this study is that the models are dependent on the accuracy of the historical queue length estimations (i.e., ground truth). From literature review, it has been observed that accurate queue length estimation is not a trivial task, so this work needs to be combined with a queue length estimation framework for effective utilization. Future work should also include the following: (1) mid-term and long-term forecasts (2) modifications to the basic Grey systems equations and a study on applicability of multivariable Grey models, and (3) seasonal behavior inducing model structures. 
    
\section*{Acknowledgments}
This study is based on a study supported by the Center for Connected Multimodal Mobility ($C^2$$M^2$) (USDOT Tier 1 University Transportation Center) Grant headquartered at Clemson University, Clemson, South Carolina, USA. The authors would also like to acknowledge U.S. Department of Homeland Security (DHS) Summer Research Team Program Follow-On, and National Science Foundation (NSF, No. 1719501) grants. Any opinions, findings, conclusions or recommendations expressed in this material are those of the author(s) and do not necessarily reflect the views of ($C^2$$M^2$), USDOT, DHS, or NSF and the U.S. Government assumes no liability for the contents or use thereof. 

\bibliographystyle{elsarticle-harv}
\bibliography{detection_trb}

\end{document}